\documentclass[amsmath,twocolumn]{aastex63}

\usepackage{enumitem}
\usepackage{amsmath}
\usepackage{units}
\usepackage{apjfonts}
\usepackage{CJK}

\usepackage{amsmath}
\usepackage{comment}

\usepackage{dcolumn}
\usepackage{bm}

\usepackage{xcolor}

\newcommand{\etal}{\textit{et al}.}

\newcommand{\eg}{\textit{e}.\textit{g}.}
\newcommand{\Ne}{$^{22}$Ne}
\newcommand{\Fe}{$^{56}$Fe}

\shorttitle{Fe Inner Cores in WDs}
\shortauthors{Caplan \etal}
\graphicspath{{./}{figures/}}

\begin{document}

\title{Cooling Delays from Iron Sedimentation and Iron Inner Cores in White Dwarfs} 

\author{M. E. Caplan}
 \email{mecapl1@ilstu.edu}
 \author{I. F. Freeman}
\affiliation{
 Illinois State University, Department of Physics, Normal, IL 61790 
}
\author{C. J. Horowitz}
\affiliation{Center for Exploration of Energy and Matter and
                  Department of Physics, Indiana University,
                  Bloomington, IN 47405, USA}
\author[0000-0002-6335-0169]{A. Cumming}
\affiliation{Department of Physics and McGill Space Institute, McGill University,
Montreal, QC H3A 2T8, Canada}

\author[0000-0003-4456-4863]{E. P. Bellinger}
\affiliation{Stellar Astrophysics Centre, Department of Physics and Astronomy, Aarhus University, Ny Munkegade 120, Aarhus, Denmark}




\begin{abstract}
Do white dwarfs have inner cores made of iron? Neutron rich nuclei like \Fe\ experience a net gravitational force and sediment toward the core. Using new phase diagrams and molecular dynamics simulations, we show that \Fe\ should separate into mesoscopic Fe-rich crystallites due to its large charge relative to the background. \deleted{At solar abundances, these crystallites rapidly precipitate, producing early time cooling delays, and form an inner core of order 100 km and $10^{-3} M_\odot$, which may be detectable with asteroseismology.}\added{At solar abundances, these crystallites rapidly precipitate and form an inner core of order 100 km and $10^{-3} M_\odot$ that may be detectable with asteroseismology. Associated cooling delays could be up to a Gyr for low mass white dwarfs but are only $\sim$0.1 Gyr for massive white dwarfs, so while this mechanism may contribute to the Q-branch the heating is insufficient to fully explain it.}
\end{abstract}




\section{Introduction} \label{sec:intro}

Recent observations of Galactic white dwarfs (WD) with Gaia, such as those resolving the latent heat released by core crystallization and the discovery of the Q-branch, have renewed interest in the physics of core crystallization and sedimentation of neutron rich nuclei  
\citep{Cheng_2019,crystallization}. 
While \Ne\ (mass fraction $X_{^{22} \rm Ne}\approx 0.02$) is the dominant sedimentary heat source, \Fe\ ($X_{^{56} \rm Fe}\approx 10^{-3}$) may be important at early times, as the greater neutron excess gives twice the heating per nucleus and a faster sedimentation timescale \citep{Isern1991,bildsten2001gravitational}. 

At a fixed electron density, nuclei with higher charges experience stronger Coulomb interactions and thus generally separate and crystallize first in a mixture. For example, \cite{PhysRevLett.126.131101} argue that actinides may separate into microgram-scale crystallites in the cores of WDs at roughly twice the C/O crystallization temperature. 
Consider the coupling parameter $\Gamma_i = e^2 Z_i^2 / a_i k_{B}T$ as an effective temperature (nuclear charge $eZ_i$, Wigner-Seitz radius $a_i = (3 Z_i / 4 \pi n_e)^{1/3}$ with electron number density $n_e$, and temperature $k_{B}T$).
Iron, with $\Gamma_\mathrm{Fe}/\Gamma_\mathrm{C} = (26/6)^{5/3} = 11.5$ will be strongly supercooled in the C/O background long before the onset of C/O crystallization.

Recently, \cite{Bauer2020} showed that single particle diffusion of \Ne\ is insufficient to produce the observed Q-branch heating assuming solar metallicity, but \Ne\ `clusters' 
of $10^2$ to $10^3$ nuclei can enhance the sedimentation rate to appropriate timescales. However, \cite{Caplan2020APJL} showed that \Ne\ does not strongly separate from the C/O background, as only nuclei with a larger charge (relative to the background) can strongly separate due to the entropy of mixing. \added{Recent work has also suggested larger \Ne\ abundances \citep{Camisassa2020} or distillation \citep{BlouinDistillation} as possible solutions.} Nevertheless, \cite{Bauer2020} laid the framework to seriously consider precipitates and their sizes in WDs, which we are motivated to consider in more detail in this work.  

Existing phase diagrams suggest that strong eutectic separation occurs in binary mixtures with charges $Z_2 / Z_1 \gtrsim 2$, with two solid phases available: (1) a nearly pure phase of $Z_2$ nuclei, and (2) an alloy of $Z_1$ and $Z_2$ nuclei which is enhanced in $Z_2$ nuclei relative to the background \citep{Ogata1993,Segretain93,Medin2010,Medin2011}. As $Z_\mathrm{Fe}/Z_\mathrm{C}= 4.3$, it is clear that we should expect Fe to separate despite its low number abundance. Crystallites, once formed, grow quickly as Fe nuclei from the background adsorb onto the surface. While such crystallites have enhanced sedimentation \citep{Bauer2020}, they may also encounter each other and combine to form aggregations which rapidly precipitate to the core. Thus, sedimentation may proceed quickly after the onset of crystallization. 

\deleted{While \Fe\ sedimentation is known in the literature, it has received little attention \citep{Segretain93,bildsten2001gravitational} and is not included in recent sedimentary models of cooling \citep{Bauer2020}. 
Nevertheless, the sedimentation of even a fraction of the $10^{-3} M_\odot$ \Fe\ (assuming solar abundances) in a $1 M_\odot$ WD can produce heating of order $10^{-4}$ to $10^{-3} L_\odot$ for a Gyr, depending on the exact sedimentation timescale, and must be included to accurately model the cooling delay.}\added{\cite{Xu1992} suggested that iron sedimentation could produce significant cooling delays, and some attention to Fe was also given in \cite{Segretain93,bildsten2001gravitational}, which now motivates work studying Fe in multi-component mixtures. The sedimentation of even a fraction of the $10^{-3} M_\odot$ \Fe\ (assuming solar abundances) in a $1 M_\odot$ WD can produce heating of order $10^{-4}$ to $10^{-3} L_\odot$ for a Gyr, depending on the exact sedimentation timescale, and must be included to accurately model the cooling delay.}

In this work we consider the precipitation of \Fe\ crystals, and show that they will reach mesoscopic sizes, collecting in the center of the star to form an Fe core of order 100 km and $10^{-3} M_\odot$, with broad implications for cosmochronology (from the cooling delay from sedimentary heating) and asteroseismology (from the stratification of the WD).

While we consider only the separation of Fe from C/O mixtures, given the large charge ratio the discussion generalizes to higher mass O/Ne/Mg WDs. In sec. \ref{sec:pd} we present the ternary C/O/Fe phase diagram which we verify with molecular dynamics in sec. \ref{sec:md}. We discuss the implications for WD structure in secs. \ref{sec:dis} and \ref{sec:sum}.

\section{C/O/Fe phase diagram}\label{sec:pd}

We begin by calculating a ternary phase diagram to determine what Fe-alloys coexist with a C/O liquid. We use the method of \cite{Medin2010}, which has been used extensively to predict the separation of ternary mixtures in our past work \citep[code available online\footnote{\url{https://github.com/andrewcumming/phase_diagram_3CP}}]{Caplan2018,Caplan2020APJL}. This method is based on analytic fits to the free-energies of mixtures \citep{Ogata1993}, and uses the double tangent construction, which identifies points on the minimum free energy surfaces that share a tangent plane.  

\begin{figure}[t!]
\centering
\includegraphics[trim=0 0 0 0,clip,width=0.49\textwidth]{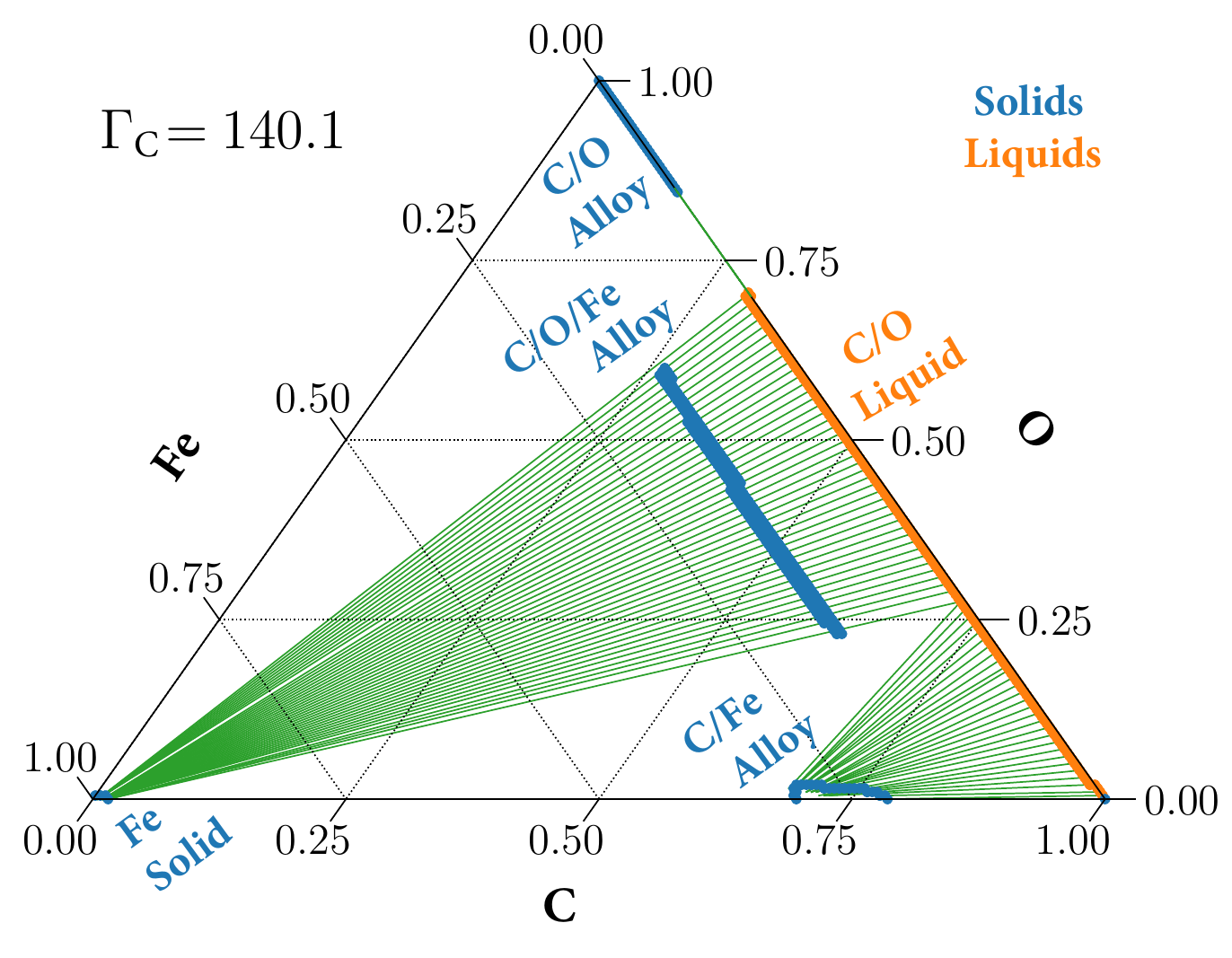}
\caption{\label{fig:PD} 
(Color online) C/O/Fe phase diagram. We show liquidus (orange) and solidus (blue) curves connected by tie-lines (green) showing coexistence.  
The right side axis is C/O mixture with trace Fe, while the bottom left corner is pure Fe. 
Pairs of points correspond to the compositions of a liquid and solid that coexist, while triplets correspond to a liquid and two solids that coexist (as in a eutectic point). We label compositions by $\vec{x} = (x_\mathrm{C}, x_\mathrm{O}, x_\mathrm{Fe})$ and we project lines of constant $x_i$ from the tick marks on the relevant axis. For example, lines parallel to the C/O axis show increasing Fe. The initial composition in a WD will be near the C/O axis.}
\end{figure}

In Fig. \ref{fig:PD} we show a C/O/Fe phase diagram at one temperature, reported in units of the carbon $\Gamma_\mathrm{C} \propto 1/T$ $(\Gamma_\mathrm{O} = 1.6 \Gamma_\mathrm{C}$, $\Gamma_\mathrm{Fe} = 11.5\Gamma_\mathrm{C}$). 
The composition of the liquid WD can be found in the orange region on the right side, corresponding to a mixture of C/O with trace Fe. 

On the C/O axis we observe the expected behavior for C/O separation in the absence of Fe, as liquid $\vec{x}_l \approx (0.3,~0.7,~0.0)$ coexists with solid $\vec{x}_s \approx (0.15,~ 0.85,~ 0.0)$, roughly consistent with the two-component C/O phase diagrams from \cite{Medin2010} and \cite{Blouin_2020}.
While only the most O-rich C/O mixtures freeze at this temperature, at higher $x_\mathrm{C}$ we find three Fe-rich alloys in coexistence with C/O liquids.  

Mixtures with comparable amounts of C and O show coexistence between a C/O liquid with trace Fe and two solid phases. As expected, the charge ratio between the Fe and C/O is between about 3 and 4, depending on the C/O ratio, and such a mixture is known to strongly separate \citep{Caplan2018}. The `island' with Fe abundances near 15\% is analogous to the `island' seen in the two-component phase diagrams in Fig. 1 of \cite{Medin2011}, which uses $Z_2 / Z_1 = 4.25$. Note that by analogy with the two-component system the `island' for the C/O/Fe alloy is actually a loop and so each coexistence line intersects two similar alloys.
At $x_\mathrm{C} \gtrsim 0.75$, we also observe the formation of a roughly $\vec{x}_s \approx (0.75, ~0.00, ~0.25)$ C/Fe alloy that is depleted in O and does not coexist with the pure Fe.

For a realistic WD with $x_\mathrm{C}/x_\mathrm{O} \sim 1$ and trace Fe, this suggests that two Fe-enhanced solid phases can form in equilibrium with the liquid in the core well before the background begins to crystallize: (1) a pure Fe solid and (2) a 15\% Fe alloy. This is in stark contrast with $^{22}$Ne, which does not strongly separate from C/O when $x_\mathrm{Ne} \lesssim 0.30$ \citep{Caplan2020APJL}.

The phase diagrams are agnostic about which of these two phases may nucleate as the fluid cools. 
It is not obvious which is more likely; nucleation of crystals in mixtures is an interesting question and should be explored in future work. While the C/O/Fe alloy has abundant C and O to draw from the liquid, their higher mobility at such low $\Gamma_\mathrm{C,O}$ may inhibit growth. Furthermore, the C/O/Fe may need to form a more complicated lattice structure than a simple body-centered cubic (bcc) and may favor very specific ratios of components; \eg\ \cite{Engstrom} predicts a hexagonal FeO$_3$C$_2$ crystal is stable. For the pure Fe phase to nucleate it may require a rare thermal fluctuation where a large number of Fe come together, though this number may be small given the high $\Gamma_\mathrm{Fe}$.   
In either case, the solid phase has a neutron excess relative to the background and will sink. While the discussion that follows focuses on the pure Fe, a C/O/Fe alloy with an equivalent number of Fe nuclei would be roughly twice as massive but produce the same sedimentary heating.

\added{Due to the large charge ratios and low Fe abundances, some free energies used to compute Fig. \ref{fig:PD} are extrapolated beyond the range they were originally fit. In the next section we use molecular dynamics (MD) simulations to verify the phase separation and the stability of pure Fe solids in a C/O background.}

\section{Molecular Dynamics}\label{sec:md}

\deleted{We now use molecular dynamics (MD) simulations to verify the phase separation and the stability of pure Fe solids in a C/O background. The alloys will be the subject of future work.} 
Our \added{MD} method is the same as in \cite{Caplan2018}. Nuclei are point particles with separation $r_{ij}$ in a periodic cubic volume interacting through a Coulomb potential $V_{ij}(r_{ij})= (e^2 Z_i Z_j / r_{ij}) e^{-r_{ij}/\lambda}$ with electron screening length $\lambda^{-1}=2\alpha^{1/2}(3\pi^2n_e)^{1/3}/\pi^{1/2}$ evolved using velocity Verlet. 

\begin{figure}[t!]
\centering
\includegraphics[trim=55 145 55 145,clip,width=0.2340\textwidth]{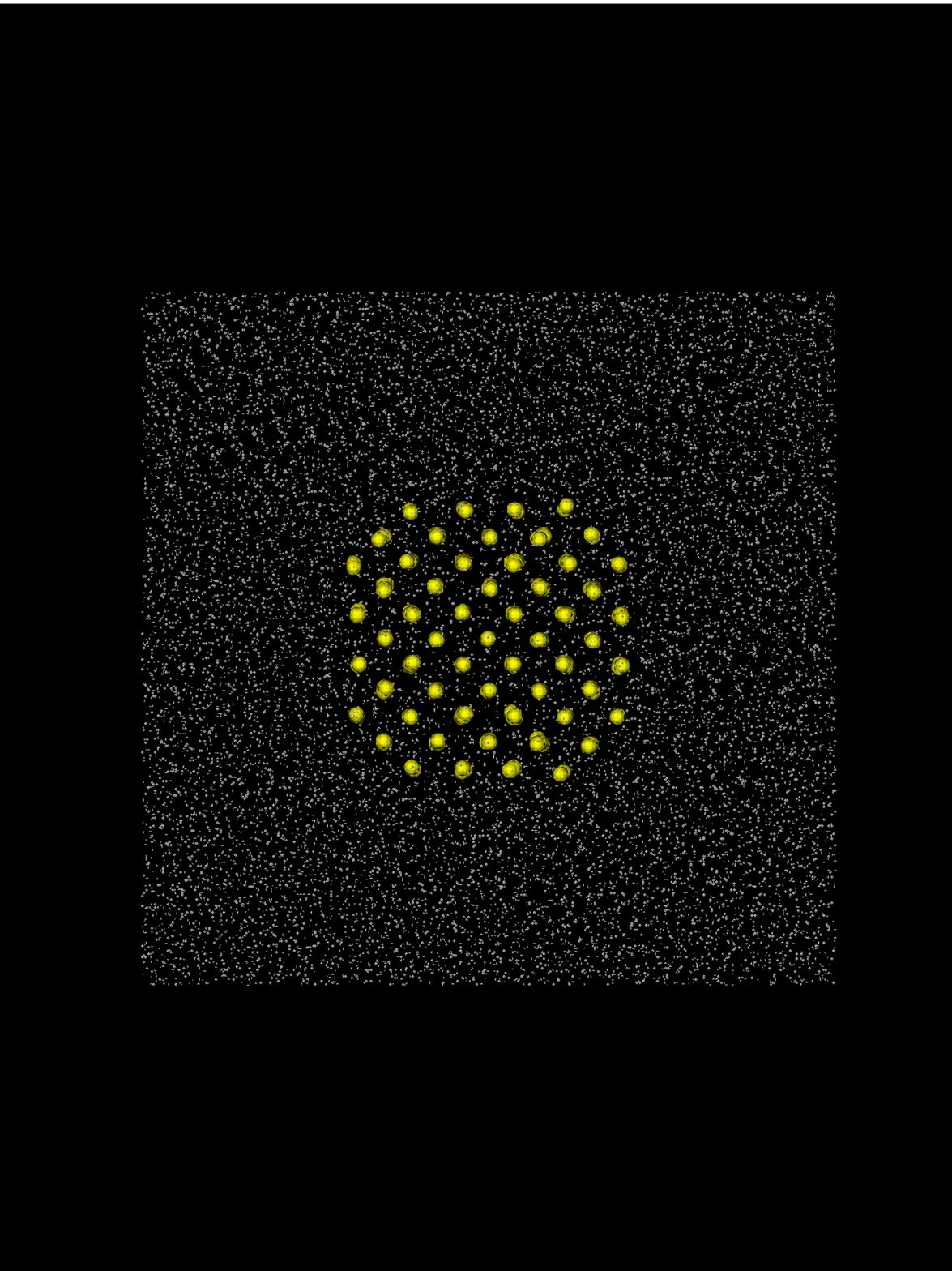}
\includegraphics[trim=55 145 55 145,clip,width=0.2340\textwidth]{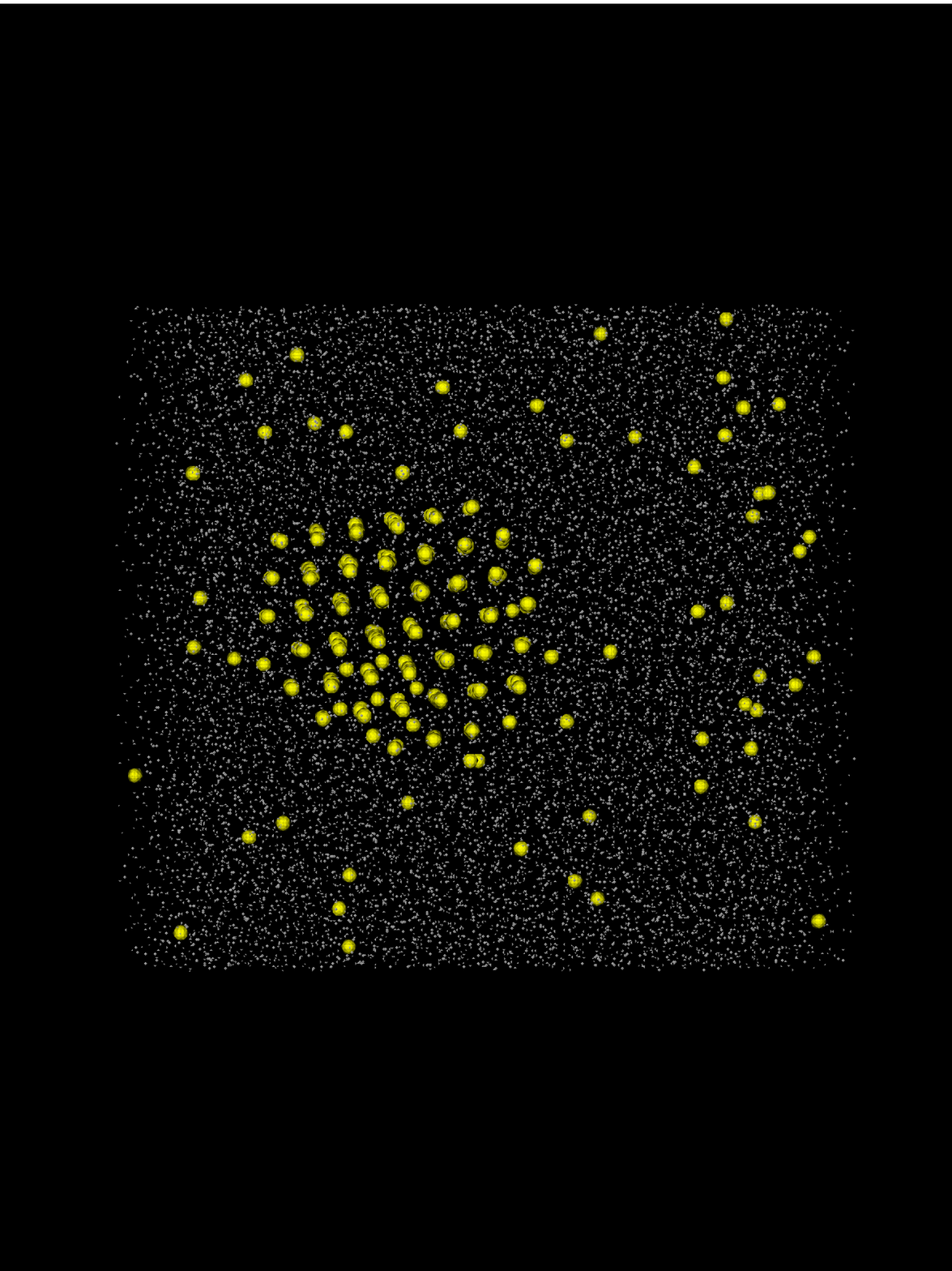}
\caption{\label{fig:md} Initial (left) and final (right) configurations for run \#5. Large points show Fe nuclei (yellow) in a C/O fluid (white).} 
\end{figure}

We report on MD simulations with $x_\mathrm{Fe} = 0.015$ (with $N=16384$ nuclei) and $x_\mathrm{Fe} = 0.0013$ ($N=65536$). As nucleation can take a long time, we begin with all of the Fe in a bcc lattice surrounded by equal amounts of C and O, as in \cite{Caplan2018} and Fig. \ref{fig:md}. When evolved, nuclei desorb from the crystal into the gas until equilibrium concentrations are found. As MD runtimes scale with $N^2$ it is difficult to run simulations with realistic Fe abundances ($x_\mathrm{Fe} \approx 10^{-4}$) for sufficiently many timesteps to reliably equilibrate.  

Simulations were evolved at constant temperature until either the crystal melted (indicating $T>T_\mathrm{melt}$), or until the size of the crystal remained constant for $10^7$ MD timesteps (indicating equilibrium). The size of the crystal was determined with a cluster algorithm and the ratio of the number of Fe in the crystal to the total number of Fe in the simulation allows us to estimate the fraction that precipitates $f_p$. The cluster size fluctuates due to stochastic adsorption and desorption; $f_p$ has about $\pm 0.02$ uncertainty. Animations of these simulations are available in the supplemental materials.\footnote{Supplemental material for this preprint available online at \\  \url{https://www.phy.ilstu.edu/~mcaplan/iron-wd-cores/}.} 
Table \ref{tab:md} summarizes our results. 

\begin{table}[]
\begin{tabular*}{0.47\textwidth}{c  @{\extracolsep{\fill}}c c c} 
\hline \hline
ID & $\vec{x} = (x_\mathrm{C}, x_\mathrm{O}, x_\mathrm{Fe})$ & $\Gamma_\mathrm{C}$  & $f_p$ \\ \hline
1  & (0.499, 0.499, 0.001)    & 183          & 0.63                             \\
2  & (0.499, 0.499, 0.001)    & 166          & 0.62                             \\
3  & (0.499, 0.499, 0.001)    & 152          &    0.00                              \\
4  & (0.499, 0.499, 0.001)    & 141          & 0.00                                \\ \hline
5  & (0.492, 0.492, 0.015)    & 185          & 0.77                                 \\
6  & (0.492, 0.492, 0.015)    & 168          &  0.63                                \\
7  & (0.492, 0.492, 0.015)    & 154          &   0.59                               \\
8  & (0.492, 0.492, 0.015)    & 143          & 0.00                                \\
9  & (0.738, 0.246, 0.015)      & 143          & 0.00                               \\
10 & (0.246, 0.738, 0.015)      & 143          & 0.63       \\
\hline \hline                         
\end{tabular*}
\caption{\label{tab:md}Summary of MD runs, including composition $\vec{x}$, inverse temperature $\Gamma_\mathrm{C}$, and the precipitation fraction $f_p$ which is a ratio of Fe nuclei in the crystal to total Fe.}
\end{table}

Our MD finds fair quantitative agreement with the phase diagram above. Simulations at $\Gamma_\mathrm{C} \lesssim 140$ find that the crystal melts rapidly while $150 \lesssim \Gamma_\mathrm{C} \lesssim 200$ reach an equilibrium with coexistence. While runs 3 ($\Gamma_\mathrm{C} = 152$) and 4 ($\Gamma_\mathrm{C}=141$) do melt, the crystal survived for many millions of MD timesteps as nuclei slowly desorbed from the surface; this metastability suggests these systems were only weakly superheated and only slightly above the melting temperature. The melting here could also be a consequence of finite size effects, and larger simulations with greater $N_\mathrm{Fe}$ may be stable.

Simulations at constant $\Gamma_\mathrm{C}$ varying $x_\mathrm{C}/x_\mathrm{O}$ also allow us to probe the robustness of this separation with respect to the background. While the Fe solid persists at $x_\mathrm{C}\approx0.25$, $x_\mathrm{O}\approx0.75$ (run 10) it melts quickly in a background of $x_\mathrm{C}\approx0.75$, $x_\mathrm{O}\approx0.25$ (run 9). Equivalent runs were also performed using an enhanced $x_\mathrm{Fe}=0.023$ with the same results. This is consistent with the prediction from the phase diagram that at high $x_\mathrm{C}$ the C/Fe alloy does not coexist with the pure Fe. 

In summary, the MD finds some sensitivity to the exact $\Gamma_\mathrm{C}$, $x_\mathrm{Fe}$, and $x_\mathrm{C} / x_\mathrm{O}$ which should be explored in future work. Nevertheless, these simulations show that the phase diagram above is qualitatively accurate and that solids strongly enhanced in Fe may form before conventional C/O crystallization begins.

\section{Discussion}\label{sec:dis}

\textit{Crystallite Growth:} We now calculate the characteristic size of the Fe crystallites.
Once nucleated, crystals grow through adsorption of Fe diffusing in the background.
For simplicity we again consider the pure Fe solid.
If nucleation is slow, Fe undergoing Brownian motion in the background encounters the cluster and adsorbs on a diffusive timescale given by $D_\mathrm{Fe} \approx r^2 / t$, where $D_\mathrm{Fe}$ is the single particle Fe diffusion coefficient, $r$ the size of the volume the Fe is taken from, and $t$ the growth time. Growth ceases when the cluster falls out of the fluid on a sedimentation timescale $v_{cl} = h/t$ given by the cluster sedimentation velocity $v_{cl}$ and the height it falls (\eg\ the WD radius),
\begin{equation}
    v_{cl} = 4 m_n g \langle N \rangle \frac{D_{cl}}{k_B T}
\end{equation}
with $4 m_n g$ being the net gravitational force, $D_{cl} = D_\mathrm{Fe}/\langle N \rangle^{1/3}$ the cluster diffusion coefficient, and $k_B T$ the temperature \citep{Bauer2020}. Equating timescales, we find the cluster size is independent of $D_\mathrm{Fe}$ and the crystallites fall out with a characteristic number of Fe nuclei

\begin{equation}
\begin{aligned}
\langle N \rangle_{sp} & = 1.3 \times 10^{21} \left( \frac{h}{6000~ \mathrm{ km}}\right)^{3/4}  \left(\frac{\rho }{ 10^8 \mathrm{ g/cm}^3} \right)  \\
& \times \left( \frac{f_p}{0.5} \right)^{3/4}  \left( \frac{x_\mathrm{Fe}}{10^{-4}} \right)^{3/4}  \left( \frac{g}{10^9~ \mathrm{ cm/s}^2}\right)^{-3/4}  \left( \frac{\Gamma_\mathrm{C} }{ 140} \right)^{-3/4} 
\end{aligned}
\end{equation}

\noindent with a density $\rho$ at height of formation $h$ with precipitation fraction $f_p$, using typical scales for a solar mass WD. 

If nucleation is fast and many small crystallites form simultaneously then they instead grow by aggregation. Aggregations grow more slowly because diffusion of crystallites is slower than single particle diffusion, so their size is calculated using $D_{cl} \approx r^2/t$, 

\begin{equation}
\begin{aligned}
\langle N \rangle_{ag} & = 7.9 \times 10^{16} \left( \frac{h}{6000~ \mathrm{  km}}\right)^{3/5}  \left(\frac{\rho }{ 10^8 \mathrm{ g/cm}^3} \right)^{4/5}  \\
& \times \left( \frac{f_p}{0.5} \right)^{3/5}  \left( \frac{x_\mathrm{Fe}}{10^{-4}} \right)^{3/5}  \left( \frac{g}{10^9~ \mathrm{ cm/s}^2}\right)^{-3/5} \left( \frac{\Gamma_\mathrm{C} }{ 140} \right)^{-3/5}. 
\end{aligned}
\end{equation}

\noindent If the timescales for nucleation and precipitation are comparable, we may expect clusters with between $10^{16}$ and $10^{21}$ nuclei which deplete their surroundings of Fe through single-particle diffusion at early times and aggregate later.

These are only rough order-of-magnitude scales. 
Regardless, it is clear that growth proceeds quickly and mesoscopic clusters with masses between $10^{-5}$ and $10^{-2}$ grams that sink with speeds between a few cm/s and a few m/s are typical. This precipitation timescale is fast; while single-particle sedimentation takes order gigayears, precipitation is faster by $\langle N \rangle^{2/3}$ and has timescales of days to years, depending on cluster sizes. These clusters accumulate in the core forming an inner core of Fe or a C/O/Fe alloy.

\textit{Core mass and radius:} We can estimate the size of this inner core by determining what volume of the star will have precipitated its iron when traditional C/O crystallization begins. This will depend on the exact C/O ratio; using the \cite{Blouin_2020} C/O phase diagram, we can expect a $x_\mathrm{C} = x_\mathrm{O}$ mixture to begin crystallizing at $\Gamma_\mathrm{C} \approx  1.1 \Gamma_{crit} \approx 190$. Given that Fe precipitation occurs at $\Gamma_\mathrm{C} \approx 140$, a core density of $\rho_c = 10^8 \mathrm{ g/cm}^3$ at $\Gamma_\mathrm{C}\approx 190$ (assuming an isothermal WD) suggests that precipitation has occurred out to densities of $4\times10^7 \mathrm{ g/cm}^3$. This density is found 1450 km above the core in a $M=1.17 M_\odot$ WD, coincidentally at a third the radius of the star and containing a third of the mass of the star. Assuming solar metallicity ($m_\mathrm{Fe} \approx 10^{-3} M_\odot$) and precipitation fraction $f_p \approx 0.5$, we find an inner core mass of $2\times10^{-4} M_\odot$ if pure Fe precipitates. At average densities of $\rho_c$, this inner core is approximately 150 km in diameter. If an alloy of 15-20\% Fe forms, we may expect an inner core two to three times more massive, depending on the exact composition. 

\begin{figure}[t!]
\centering
\includegraphics[trim=0 250 220 50,clip,width=0.49\textwidth]{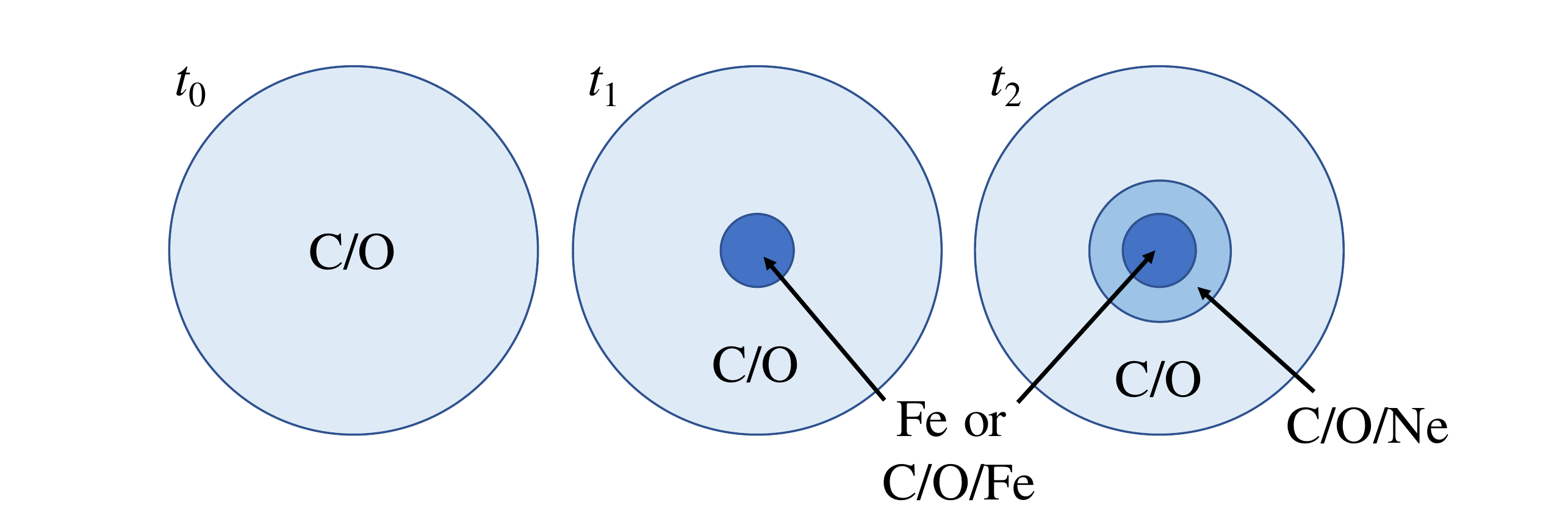}
\caption{\label{fig:core} Schematic illustration of the time evolution of the WD from formation (left), to Fe precipitation (center), through Ne shell formation (right). Not to scale.} 
\end{figure}

This estimate may be a lower limit as other processes may transport more Fe into the region where it precipitates before the background freezes and stalls the growth of the inner core. For example, as $\Gamma_\mathrm{C}$ increases $f_p$ may increase, so clusters may continue to grow while falling through partially depleted regions. Single-particle sedimentation from the upper layers of the star will also enhance the core Fe abundance, which should be possible to model with stellar evolution codes such as MESA. \deleted{Furthermore, if the first solids to form when the background begins to freeze tend to expel neutron-rich nuclei such as \Fe\ or \Ne\, those crystals will be buoyant relative to the surrounding liquid and initiate the distillation process of \cite{BlouinDistillation}}\added{The Fe depleted fluid, having a higher $Y_e$, may also be buoyant and mix with the outer layers which could maintain a homogeneous liquid composition throughout the star \citep{Xu1992}}. This may further enhance the abundance of \Fe\ in the core, as well as \Ne\ which might be expected to form a shell as a sort of `outer core' when it begins to crystallize (Fig. \ref{fig:core}).  

The onset of C/O crystallization around the Fe core (or, more likely, a Ne-alloy shell), does not necessarily stop the sedimentation process, as Fe precipitation will proceed above. For the purposes of estimating total heating, one should consider the entire Fe content of the star.
These crystallites may produce mesoscopic inclusions in whatever crystal surrounds the Fe inner core. It is traditionally assumed that the composition is frozen in once crystallization occurs, but \cite{Hughto2011} finds single-particle diffusion coefficients are only two to three orders of magnitude lower in crystals at $\Gamma \approx 200$ than in the strongly-coupled liquid. Even though diffusion in the lattice is exponentially suppressed with temperature, given the fast timescales for sedimentation calculated above it is possible that some of these inclusions could migrate through the shell to reach the Fe core before this process is quenched. Viscoelastic creep and diffusion at crystal grain boundaries is an interesting question which could be studied with MD. Though beyond the scope of this work, assumptions about the behavior of solid phases in WDs should be revisited (see also \cite{Mckinven2016} sec. 3.1).

\textit{Cooling Delay:} 
Precipitation may release \replaced{a few times}{of order} $10^{46}$ ergs\added{, as in \cite{Xu1992},} while forming an inner core of $M\lesssim 10^{-3} M_\odot$. 
This is obviously sensitive to the metallicity and mass of the WD, so we consider this a rough energy scale. Latent heat from freezing may be comparable to the sedimentation energy and scales with the crystallized mass, and thus is larger for alloys. 

We estimate the cooling delay from the luminosity of the WD at the onset of precipitation. The $M=0.52 \ M_\odot$ ($Z=0.01$) cooling model of \cite{Renedo2010} reaches core $\Gamma_\mathrm{C}=140$ at $\log_{10}(L/L_\odot)=-3.9$ after 3.3 Gyr. At this luminosity we expect a delay of order 1 Gyr from Fe precipitation prior to traditional C/O crystallization; the luminosity does not change significantly over the course of precipitation ($\log_{10}(L/L_\odot)=-4.1$ at core $\Gamma_\mathrm{C} = 200$). Their $M=0.93 \ M_\odot$ ($Z=0.01$) cooling model reaches core $\Gamma_\mathrm{C}=140$ at $\log_{10}(L/L_\odot)=-2.9$ in 0.9 Gyr and may produce an order 0.1 Gyr cooling delay. Though perhaps an order of magnitude smaller than the \Ne\ delay, this mechanism is rapid and the heat release is large at early times and may be important to include for precision cosmochronology.

\section{Summary}\label{sec:sum}

Mesoscopic crystallites of Fe should precipitate to the center of WDs to form a macroscopic Fe inner core. This precipitation is a natural consequence of the high charge of Fe relative to C/O which causes it to separate. Solids with high Fe concentrations relative to the background form at temperatures above the C/O crystallization point. This inner core may be either a nearly pure Fe crystal or a C/O/Fe alloy, depending on the exact composition of the star and the nucleation physics.  
Because this mechanism is efficient, rapidly transporting approximately half the Fe in the star to the core, it is a powerful source of gravitational potential energy to delay cooling and should be modeled with stellar evolution codes like MESA \citep{Paxton_2019,Bauer2020}.

An Fe core could be detectable with asteroseismology. 
WDs have been observed for decades to pulsate with internal gravity waves ($g$ modes) of low radial order that are roughly evenly spaced in period \citep[see, e.g.,][for a review]{2020FrASS...7...47C}. 
Precise measurements of these periods have been used to estimate their mass, including the mass of their crystal cores, as well as their internal composition profiles \citep[e.g.,][]{2004ApJ...605L.133M,2018Natur.554...73G,2018ApJ...867L..30T, 2019A&A...632A.119C}. 
\citet{1999ApJ...526..976M} studied the effect of core crystallization on WD pulsations, and found that $g$ modes are unable to penetrate the crystal core. 
Consequently, the inner boundary of the oscillations moves outward as the crystal core grows, causing an increased mean period spacing with increasing crystal mass fraction. 
Therefore, at the same luminosity and effective temperature, a WD with a crystal Fe core would exhibit a markedly different oscillation spectrum than one without. 
In a similar vein, \cite{Chidester2021} recently showed that the g-mode pulsations of low-mass WDs are measurably different when a constant profile of 2\% \Ne\ is included due to the sensitivity of the equation of state to the electron fraction $Y_e$. A similar (but weaker) effect could be expected with the inclusion of Fe, which is traditionally neglected in WD models. 
Finally, transitions in the abundances between the non-crystallized zones are well-known to cause ``bumps'' in the Brunt--V\"ais\"al\"a profile, which causes modulation of $g$-mode periods \citep[e.g.,][]{1999ApJ...526..976M,Chidester2021}. 
It may even be possible to determine the composition of the background; greater background charges will decrease the Fe precipitation fraction and also the equilibrium concentrations of Fe in an alloy, and requires future work on phase diagrams of mixtures such as O/Ne/Fe. 
This motivates including sedimentation, precipitation, and modern phase diagrams in evolutionary models in order to study the core structure of these WDs. 


Fe inner cores, if present, may impact the ignition of supernova. As pure Fe does not burn, the ignition would be off-center. A C/O/Fe alloy meanwhile may have abundant C/O available for burning at slightly higher matter densities (and screening) due to the presence of Fe and burning could easily be explored in 2D supernova codes. Similarly, precipitation increases the core density and electron Fermi energy; the most massive WDs may explode in supernova if this effect is large enough to initiate electron capture reactions \citep{caplan2020black,caiazzo2021highly}.

This mechanism also generalizes to other high $Z$ nuclei. Despite their low abundance, some high $Z$ nuclei may separate and precipitate (such as the actinide crystallites considered in \cite{PhysRevLett.126.131101}). Though \Fe\ and \Ne\ are dominant, there may be several smaller concentric shells ordered radially by decreasing charge.

\acknowledgments
CH's research was supported in part by US Department of Energy Office of Science grants DE-FG02-87ER40365 and DE-SC0018083. The authors acknowledge the Indiana University Pervasive Technology Institute for providing supercomputing and database, storage resources that have contributed to the research results reported within this paper. This research was supported in part by Lilly Endowment, Inc., through its support for the Indiana University Pervasive Technology Institute. AC is supported by an NSERC Discovery Grant, and is a member of the Centre de recherche en astrophysique du Québec (CRAQ).
Funding for the Stellar Astrophysics Centre is provided by The Danish National Research Foundation (Grant agreement no.: DNRF106).


\begin{thebibliography}{}
\expandafter\ifx\csname natexlab\endcsname\relax\def\natexlab#1{#1}\fi
\providecommand{\url}[1]{\href{#1}{#1}}

\bibitem[{{Bauer} {et~al.}(2020){Bauer}, {Schwab}, {Bildsten}, \&
  {Cheng}}]{Bauer2020}
{Bauer}, E.~B., {Schwab}, J., {Bildsten}, L., \& {Cheng}, S. 2020, \apj, 902,
  93

\bibitem[{Bildsten \& Hall(2001)}]{bildsten2001gravitational}
Bildsten, L., \& Hall, D.~M. 2001, The Astrophysical Journal Letters, 549, L219

\bibitem[{{Blouin} {et~al.}(2021){Blouin}, {Daligault}, \&
  {Saumon}}]{BlouinDistillation}
{Blouin}, S., {Daligault}, J., \& {Saumon}, D. 2021, \apjl, 911, L5

\bibitem[{Blouin {et~al.}(2020)Blouin, Daligault, Saumon, Bédard, \&
  Brassard}]{Blouin_2020}
Blouin, S., Daligault, J., Saumon, D., Bédard, A., \& Brassard, P. 2020, A\&A,
  640, L11.
\newblock \url{http://dx.doi.org/10.1051/0004-6361/202038879}

\bibitem[{Caiazzo {et~al.}(2021)Caiazzo, Burdge, Fuller, Heyl, Kulkarni,
  Prince, Richer, Schwab, Andreoni, Bellm, {et~al.}}]{caiazzo2021highly}
Caiazzo, I., Burdge, K.~B., Fuller, J., {et~al.} 2021, Nature, 595, 39

\bibitem[{{Camisassa} {et~al.}(2020){Camisassa}, {Althaus}, {Torres},
  {C{\'o}rsico}, {Cheng}, \& {Rebassa-Mansergas}}]{Camisassa2020}
{Camisassa}, M.~E., {Althaus}, L.~G., {Torres}, S., {et~al.} 2020, arXiv
  e-prints, arXiv:2008.03028

\bibitem[{Caplan(2020)}]{caplan2020black}
Caplan, M. 2020, Monthly Notices of the Royal Astronomical Society, 497, 4357

\bibitem[{Caplan {et~al.}(2018)Caplan, Cumming, Berry, Horowitz, \&
  Mckinven}]{Caplan2018}
Caplan, M.~E., Cumming, A., Berry, D.~K., Horowitz, C.~J., \& Mckinven, R.
  2018, The Astrophysical Journal, 860, 148.
\newblock \url{https://doi.org/10.3847\%2F1538-4357\%2Faac2d2}

\bibitem[{Caplan {et~al.}(2020)Caplan, Horowitz, \& Cumming}]{Caplan2020APJL}
Caplan, M.~E., Horowitz, C.~J., \& Cumming, A. 2020, The Astrophysical Journal,
  902, L44.
\newblock \url{https://doi.org/10.3847/2041-8213/abbda0}

\bibitem[{Cheng {et~al.}(2019)Cheng, Cummings, \& M{\'{e}}nard}]{Cheng_2019}
Cheng, S., Cummings, J.~D., \& M{\'{e}}nard, B. 2019, The Astrophysical
  Journal, 886, 100

\bibitem[{{Chidester} {et~al.}(2021){Chidester}, {Timmes}, {Schwab},
  {Townsend}, {Farag}, {Thoul}, {Fields}, {Bauer}, \&
  {Montgomery}}]{Chidester2021}
{Chidester}, M.~T., {Timmes}, F.~X., {Schwab}, J., {et~al.} 2021, \apj, 910, 24

\bibitem[{{C{\'o}rsico}(2020)}]{2020FrASS...7...47C}
{C{\'o}rsico}, A.~H. 2020, Frontiers in Astronomy and Space Sciences, 7, 47

\bibitem[{{C{\'o}rsico} {et~al.}(2019){C{\'o}rsico}, {De Ger{\'o}nimo},
  {Camisassa}, \& {Althaus}}]{2019A&A...632A.119C}
{C{\'o}rsico}, A.~H., {De Ger{\'o}nimo}, F.~C., {Camisassa}, M.~E., \&
  {Althaus}, L.~G. 2019, \aap, 632, A119

\bibitem[{{Engstrom} {et~al.}(2016){Engstrom}, {Yoder}, \& {Crespi}}]{Engstrom}
{Engstrom}, T.~A., {Yoder}, N.~C., \& {Crespi}, V.~H. 2016, \apj, 818, 183

\bibitem[{{Giammichele} {et~al.}(2018){Giammichele}, {Charpinet}, {Fontaine},
  {Brassard}, {Green}, {Van Grootel}, {Bergeron}, {Zong}, \&
  {Dupret}}]{2018Natur.554...73G}
{Giammichele}, N., {Charpinet}, S., {Fontaine}, G., {et~al.} 2018, \nat, 554,
  73

\bibitem[{Horowitz \& Caplan(2021)}]{PhysRevLett.126.131101}
Horowitz, C.~J., \& Caplan, M.~E. 2021, Phys. Rev. Lett., 126, 131101.
\newblock \url{https://link.aps.org/doi/10.1103/PhysRevLett.126.131101}

\bibitem[{{Hughto} {et~al.}(2011){Hughto}, {Schneider}, {Horowitz}, \&
  {Berry}}]{Hughto2011}
{Hughto}, J., {Schneider}, A.~S., {Horowitz}, C.~J., \& {Berry}, D.~K. 2011,
  \pre, 84, 016401

\bibitem[{{Isern} {et~al.}(1991){Isern}, {Hernanz}, {Mochkovitch}, \&
  {Garcia-Berro}}]{Isern1991}
{Isern}, J., {Hernanz}, M., {Mochkovitch}, R., \& {Garcia-Berro}, E. 1991,
  \aap, 241, L29

\bibitem[{{Mckinven} {et~al.}(2016){Mckinven}, {Cumming}, {Medin}, \&
  {Schatz}}]{Mckinven2016}
{Mckinven}, R., {Cumming}, A., {Medin}, Z., \& {Schatz}, H. 2016, \apj, 823,
  117

\bibitem[{{Medin} \& {Cumming}(2010)}]{Medin2010}
{Medin}, Z., \& {Cumming}, A. 2010, \pre, 81, 036107

\bibitem[{{Medin} \& {Cumming}(2011)}]{Medin2011}
---. 2011, \apj, 730, 97

\bibitem[{{Metcalfe} {et~al.}(2004){Metcalfe}, {Montgomery}, \&
  {Kanaan}}]{2004ApJ...605L.133M}
{Metcalfe}, T.~S., {Montgomery}, M.~H., \& {Kanaan}, A. 2004, \apjl, 605, L133

\bibitem[{{Montgomery} \& {Winget}(1999)}]{1999ApJ...526..976M}
{Montgomery}, M.~H., \& {Winget}, D.~E. 1999, \apj, 526, 976

\bibitem[{{Ogata} {et~al.}(1993){Ogata}, {Iyetomi}, {Ichimaru}, \& {van
  Horn}}]{Ogata1993}
{Ogata}, S., {Iyetomi}, H., {Ichimaru}, S., \& {van Horn}, H.~M. 1993, \pre,
  48, 1344

\bibitem[{Paxton {et~al.}(2019)Paxton, Smolec, Schwab, Gautschy, Bildsten,
  Cantiello, Dotter, Farmer, Goldberg, Jermyn, \& et~al.}]{Paxton_2019}
Paxton, B., Smolec, R., Schwab, J., {et~al.} 2019, The Astrophysical Journal
  Supplement Series, 243, 10.
\newblock \url{http://dx.doi.org/10.3847/1538-4365/ab2241}

\bibitem[{{Renedo} {et~al.}(2010){Renedo}, {Althaus}, {Miller Bertolami},
  {Romero}, {C{\'o}rsico}, {Rohrmann}, \& {Garc{\'\i}a-Berro}}]{Renedo2010}
{Renedo}, I., {Althaus}, L.~G., {Miller Bertolami}, M.~M., {et~al.} 2010, \apj,
  717, 183

\bibitem[{{Segretain} \& {Chabrier}(1993)}]{Segretain93}
{Segretain}, L., \& {Chabrier}, G. 1993, \aap, 271, L13

\bibitem[{{Timmes} {et~al.}(2018){Timmes}, {Townsend}, {Bauer}, {Thoul},
  {Fields}, \& {Wolf}}]{2018ApJ...867L..30T}
{Timmes}, F.~X., {Townsend}, R. H.~D., {Bauer}, E.~B., {et~al.} 2018, \apjl,
  867, L30

\bibitem[{Tremblay {et~al.}(2019)Tremblay, Fontaine, Fusillo, Dunlap,
  G{\"a}nsicke, Hollands, Hermes, Marsh, Cukanovaite, \&
  Cunningham}]{crystallization}
Tremblay, P.-E., Fontaine, G., Fusillo, N. P.~G., {et~al.} 2019, Nature, 565,
  202.
\newblock \url{https://doi.org/10.1038/s41586-018-0791-x}

\bibitem[{{Xu} \& {van Horn}(1992)}]{Xu1992}
{Xu}, Z.~W., \& {van Horn}, H.~M. 1992, \apj, 387, 662

\end{thebibliography}
\providecommand{\noopsort}[1]{}\providecommand{\singleletter}[1]{#1}%

\appendix

\section{Supplemental Materials}

\begin{figure}[h]
\centering
\includegraphics[trim=0 0 0 0,clip,width=0.66\textwidth]{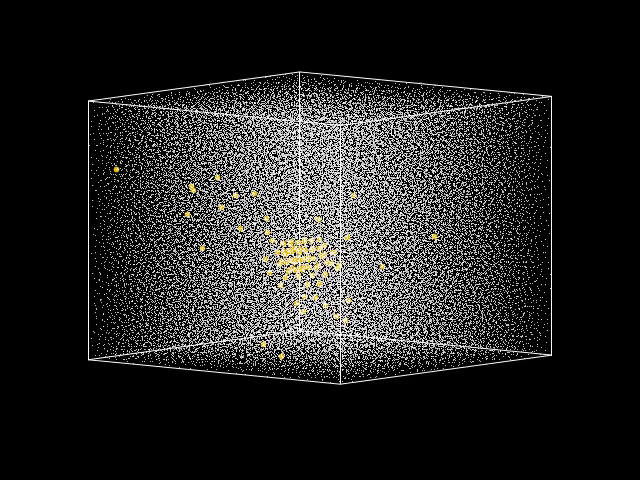}
\caption{\label{fig:SM1} 
The animation in SM1.mp4 (\href{https://www.phy.ilstu.edu/~mcaplan/iron-wd-cores/SM1.mp4}{click to download, 7.6 MB}) begins by showing the initial conditions of run \#3, with a crystal of pure iron (yellow) surrounded by a background fluid of carbon and oxygen (white). The background fades out to more clearly show the body-centered crystal lattice structure of the iron, and then fades back in before the simulation begins. When the simulation is evolved in time nuclei can be observed melting off the surface of the crystal and mixing into the background. Iron nuclei collide with carbon and oxygen in the background and undertake a random walk. The Brownian motion can be observed for both the single-particle iron nuclei in the gas and for the cluster, though diffusion of the cluster is suppressed due to its larger size. Near the end of the animation the carbon and oxygen background is again removed for clarity. This simulation contains 65,536 particles in a cubic volume with periodic boundary conditions; only 91 of them are iron, for an iron number abundance of 0.14\%. The initial conditions have the iron localized to a pure crystal, as simulating nucleation starting from diffuse Fe in a C/O background is more computationally intensive than simulating a crystal melting. In either case, we expect the equilibrium abundances in the crystal and the background to be same.}
\end{figure}

\begin{figure}[h]
\centering
\includegraphics[trim=0 0 0 0,clip,width=0.66\textwidth]{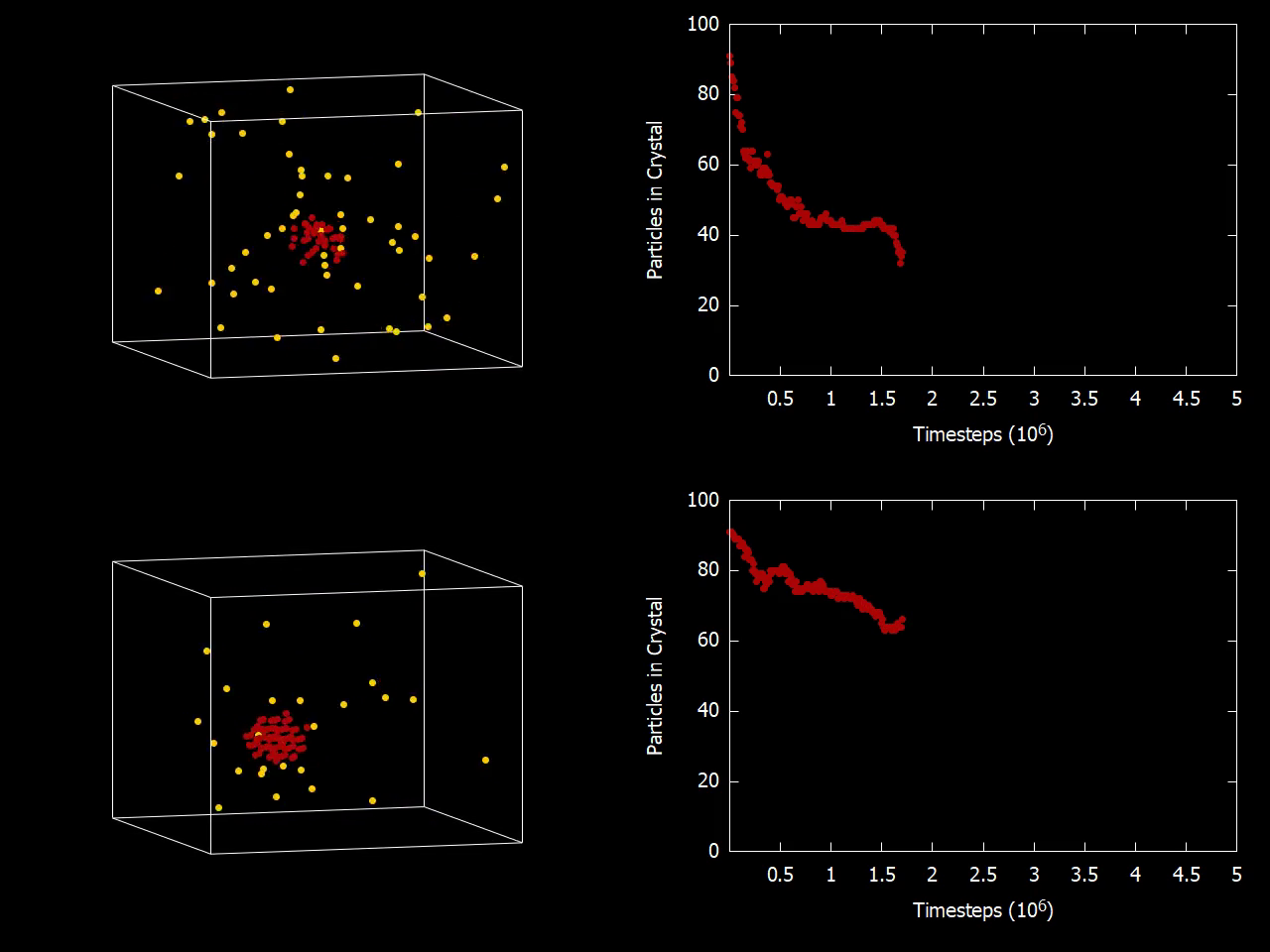}
\caption{\label{fig:SM2} The animation in SM2.mp4 (\href{https://www.phy.ilstu.edu/~mcaplan/iron-wd-cores/SM2.mp4}{click to download, 1.7 MB}) shows two simulations, run \#3 (top) and run \#1 (bottom), and the evolution of the size of their iron clusters over time. These simulations begin from identical initial conditions but have different temperatures, allowing us to study the temperature dependence of the iron phase separation in a white dwarf. We animate the simulations as in SM1, showing only the iron nuclei for clarity. Iron nuclei which the cluster algorithm identifies as members of the crystal are recolored red. In run \#3 (top, $\Gamma_\mathrm{C} = 153$) we observe that the crystal melts in approximately two million time steps after a brief period of metastability, so although this simulation is clearly above the melting temperature it is likely very close to it. In run \#1 (bottom, $\Gamma_\mathrm{C} = 183$) we see that nuclei desorb from the surface of the cluster one at a time over the first few million timesteps, but the cluster size stabilizes and finds an equilibrium iron concentration in the solid and liquid. This simulation was run for an additional ten million timesteps to verify the stability. It is also interesting to observe the slower diffusion in the colder simulation (bottom), and that during the latter half of the animation nuclei can be seen in equilibrium adsorbing and desorbing from the crystal. This gives us a percent-level uncertainty in the equilibrium iron concentrations.
}
\end{figure}



\end{document}